\documentclass[%
superscriptaddress,
frontmatterverbose, 
bibnotes,
 amsmath,amssymb,
 aps,
floatfix,
]{revtex4-2}

\usepackage{graphicx}
\usepackage{dcolumn}
\usepackage{bm}
\usepackage{hyperref}
\usepackage{graphicx}
\usepackage{epstopdf, epsfig}
\usepackage[percent]{overpic}

\usepackage{blindtext}
\usepackage{graphicx}
\usepackage{float}
\usepackage{bm}
\usepackage{multirow}
\usepackage{xcolor}

\usepackage[utf8]{inputenc}
\usepackage{xcolor}
\usepackage{float}
\usepackage{caption}
\usepackage{subcaption}

\begin{document}

\preprint{APS/123-QED}

\title{To roll or not to roll(?) is the yield stress (in soft particulate gels)}
\thanks{A footnote to the article title}%

\author{Rishabh V. More}
 \email{rishabh.more@monash.edu}
 \affiliation{Department of Chemical Engineering, Monash University, Clayton, Australia}
 \affiliation{%
Department of Mechanical Engineering, Massachusetts Institute of Technology,
Cambridge, MA 02139, USA
}
\author{Gareth H. McKinley}%
\affiliation{%
Department of Mechanical Engineering, Massachusetts Institute of Technology,
Cambridge, MA 02139, USA
}%




\date{\today}

\begin{abstract}
While it is widely acknowledged that system-spanning particulate structures contribute to the observed yield stress and shear-thinning in attractive colloidal gels, a comprehensive understanding of the underlying microscopic mechanisms remains elusive. In this study, we present findings from coarse-grained simulations focusing on model depletion gels to shed light on this intriguing phenomenon. Contrary to conventional belief, our simulations reveal that the mere presence of attractive interactions and aggregate formation does not sufficiently explain the observed yield stress. Instead, we identify a crucial physics element in the form of microscopic constraints on the relative rotational motion between bonded particles. Through a detailed analysis of microstructure and particle dynamics, we elucidate how these constraints lead to the emergence of yield stress in soft particulate gels. This research provides essential insights into the micromechanical origins of yield stress in soft particulate gels, paving the way for improved understanding and engineering of these versatile materials for various real-world applications.
\end{abstract}

\maketitle


\textit{Introduction.}---Colloidal dispersions are a versatile class of materials with wide-ranging applications across industries, including consumables, pharmaceuticals, cosmetics, and tissue engineering \cite{varga2015hydrodynamics, hilali2022sheared}. Characterized by a dispersed phase of colloidal particles in a fluid medium, these gels exhibit complex rheology driven by various microscopic interactions of a reversible or irreversible nature. Especially, they form an aggregated network through attractive interactions such as Van der Waals or depletion forces \cite{varadan2001shear, lu2013colloidal}. The resulting quiescent structures and their evolution in time have been studied extensively with respect to governing parameters such as particulate volume fraction $\phi$, the strength of inter-particle attractive potential $U_A$, and its range $\delta$ \cite{lu2013colloidal,varga2015hydrodynamics}. 

At intermediate volume fractions $10 \% \leq \phi \leq 50 \%$ and moderate reversible attractions, the networks are found to be in an arrested glassy state at small length scales to domain-spanning network at long length scales \cite{zaccone2009elasticity}. Elasticity emerges in these systems as the interconnections between minimal glassy, load-bearing clusters increase with the strength of attraction and time \cite{whitaker2019colloidal}, and has been fairly well understood \cite{bantawa2023hidden}. Notably, this structure \textit{build-up} with time from an initial disordered homogeneous state results in the aging of colloidal dispersions \cite{abou2001aging}. Under a finite deformation, these aggregated networks undergo gradual rearrangements with progressively increasing load \cite{massaro2020viscoelastic, jamali2017microstructural}. These rearrangements and progressive \textit{destruction} of the structure have been believed to result in yielding at a critical finite stress known as the yield stress $\sigma_y$, beyond which flow occurs \cite{bonn2017yield}. Upon yielding, the material exhibits shear-thinning rheology in a steady state and thixotropy, i.e., time-dependent behavior \cite{larson2019review}. 

The success of the studies as mentioned above and Brownian dynamics simulations of model attractive colloids at rest in the intermediate volume fraction regime has created an observational bias, which only considers attraction-induced aggregation as a possible cause of yield stress in soft particulate gels \cite{bonacci2022yield}. However, the microscopic mechanism for the emergence of yield stress in highly dilute attractive soft particulate gels remains elusive. Highly dilute colloidal dispersions ($\phi \leq 10 \%$), which are below the critical volume fraction of percolation \cite{campbell2005dynamical}, a percolated network structure is not possible, i.e., these systems remain weakly flocculated \cite{dullaert2005model}. As a result, attraction-induced clustering alone is insufficient to give rise to yield stress as there may not be enough interconnections (or \textit{constraints} on particle motion) between load-bearing glassy clusters to sustain finite stress at low deformations. Furthermore, clusters may not be dynamically arrested in dilute regime \cite{zhang2024percolation}. Despite this fact, highly dilute colloidal dispersions such as gels of sub-micron particles in microgravity ($\phi = 2 \%$) \cite{manley2005time} and the model \textit{thixotropic} colloidal dispersion of fumed silica ($\phi = 2.9 \%$) \cite{dullaert2005model} have been shown to exhibit finite yield stress. Thus, micro-mechanical underpinnings of yield stress in dilute colloidal dispersions hitherto are not completely understood.

\begin{figure*}
\includegraphics[width=1\textwidth]{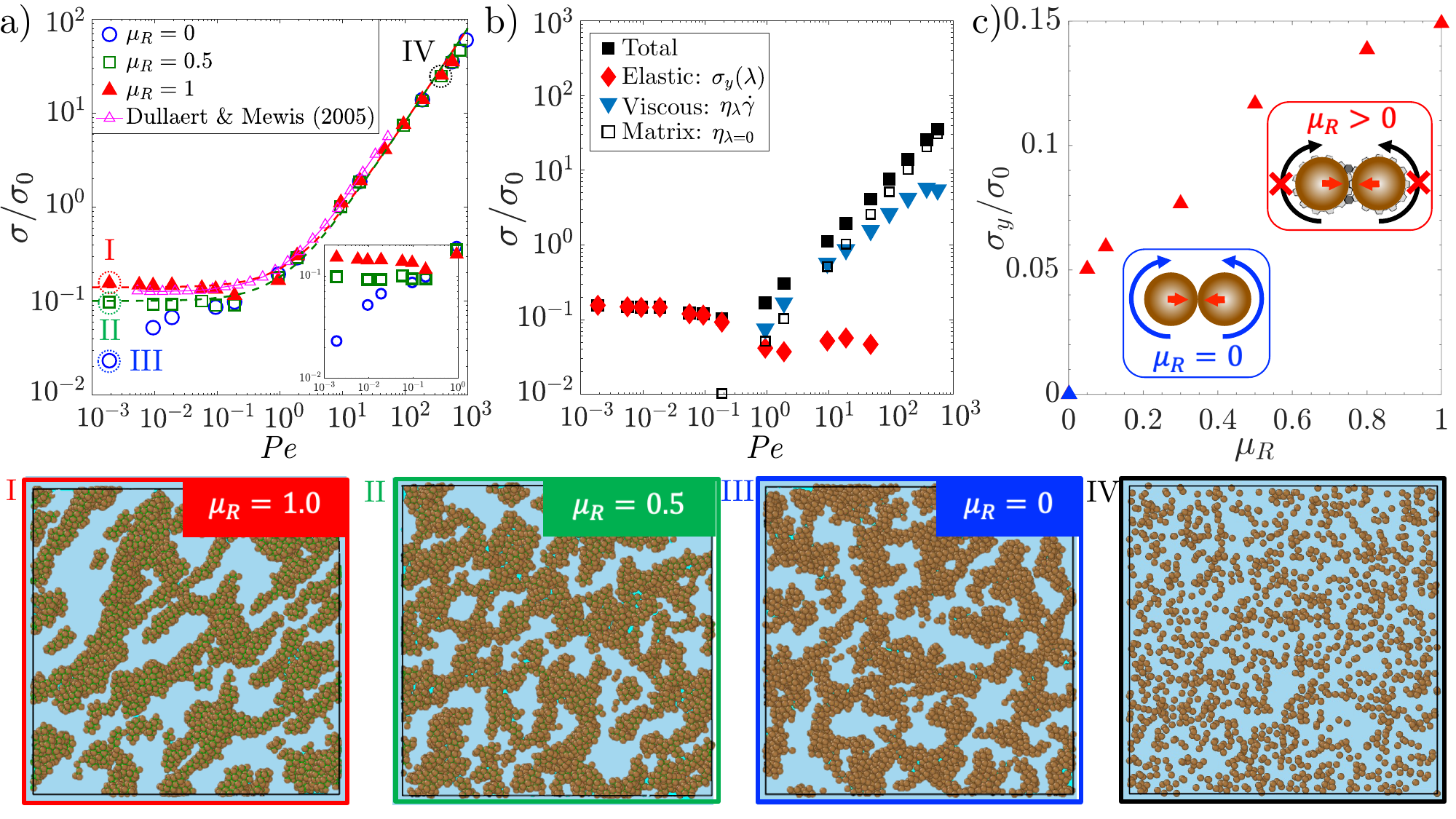}
\caption{\label{fig:fig1}a) Steady-state flow curves of $\phi=5 \, \%$ attractive depletion gels for three rolling friction values with the same attraction strengths. The simulation results agree well with the steady state flow curve of a model thixotropic fluid \cite{dullaert2005model} with $\sigma_0=30$ Pa and $\dot{\gamma}_0=2 \, s^{-1}$ for the experimental data. 
b) Various structure parameter $\lambda$ and shear-rate $\dot{\gamma}$ dependent contributions to the total dimensionless shear stress $\sigma$ as a function of $Pe$ for $\mu_R=1$. c) Dimensionless yield stress increases with the increase in rolling friction. Bottom row: Representative snapshots of the attractive aggregates in the shear plane in the yield region (low $Pe \approx  0.002$: I, II, III in (a)) for three different $\mu_R$. 
At $Pe \approx 200$ (IV), the structure is destroyed due to high shearing. \vspace{-6mm}} 
\end{figure*}

To this end, using discrete particle simulations, we show that attraction-induced aggregation is necessary but not enough for the presence of yield stress in dilute attractive colloidal dispersions. Attractive forces, which are center-to-center, not only bring colloidal particles together to form clustered structures, but they also introduce solid-solid inter-particle contacts leading to not only center-to-center but also tangential forces and torques \cite{singh2020shear}. These result in constraints on the relative sliding and rolling motion between particles and have been shown to result in macroscopic aging in moderately dense colloidal suspensions \cite{bonacci2020contact, bonacci2022yield}. We show that the yield stress arises from the constraints on the relative rolling motion between the particles in the aggregation clusters. This results in a resistance to the purely angular velocity component of the applied shear deformatio,n manifesting as the yield stress at low shear rates. The calculated structure factor patterns reveal the distinct micro-structural signature of these rolling constraints, which could be validated using the scattering pattern in beam scattering experiments. 

\textit{Governing interactions.}---We simulate the shear flow of 8192 neutrally buoyant inertia-less spherical particles with a radius $a$ in a cubical domain for a fixed volume fraction of $\phi=5 \%$ suspended in a Newtonian fluid with a constant viscosity $\eta_0$. The imposed shear rate is $\dot{\gamma}$ with Lees-Edwards periodic boundary conditions on all the sides. The particles experience Stokes drag as well as pair-wise inter-particle interactions such as hydrodynamic lubrication $\mathbf{F}_H$ \cite{ball1997simulation}, attractive depletion forces $\mathbf{F}_A$ \cite{varga2015hydrodynamics}, solid-solid contract forces $\mathbf{F}_C$ \cite{singh2020shear}, and Brownian force $\mathbf{F}_B$ \cite{kumar2010microscale} as well as the corresponding torques $\mathbf{T}_H$, $\mathbf{T}_A$, $\mathbf{T}_C$, and $\mathbf{T}_B$. To satisfy the fluctuation-dissipation theorem, we produce Brownian forces that follow $\langle \mathbf{F}_B \, \mathbf{F}_B \rangle = (2k_BT/\Delta t) \, \mathbf{\mathcal{R}}$ and $\langle \mathbf{F}_B \rangle = \mathbf{0}$. $k_B$ is the Boltzmann constant, and $T$ is the temperature in absolute units with $k_BT$ being the thermal energy magnitude, and $\Delta t$ is an appropriately chosen simulation time-step. $\mathbf{\mathcal{R}}$ is the overall resistance tensor for the system and also gives the hydrodynamic interactions between the particles as $\mathbf{F}_H = \eta_0 \mathbf{\mathcal{R}} \cdot \begin{pmatrix} \mathbf{U}^{\infty}-\mathbf{U}\\ \mathbf{E}^{\infty} \end{pmatrix}$. The imposed linear flow field is divided into a uniform velocity vector $\mathbf{U}^{\infty}$ and a strain-rate tensor $\mathbf{E}^{\infty}$ at the particles' centres moving with velocities $\mathbf{U}$. For a simple shear flow, $\mathbf{U}^{\infty} = (\dot{\gamma}y,0,0)$ with $\dot{\gamma}$ is the constant shear rate with $\mathbf{x}$, $\mathbf{y}$, and $\mathbf{z}$ being the flow, gradient and vorticity directions, respectively. We vary the Péclet number $Pe=6\pi\eta_0a^3\dot{\gamma}/k_BT$ to change the imposed shear rate and use $\sigma_0=k_BT/a^3$ as the characteristic stress. 

One common mechanism to achieve gelation is adding non-absorbing polymer to colloidal dispersion \cite{dullaert2005model, varga2015hydrodynamics}. It promotes inter-particle attraction called depletion interaction, which can be modeled using the Asakura-Oosawa potential. The depletion attraction force as a function of particle separation $r$ acting normally is $|\mathbf{F}_A|=U_A \, (3\tilde{a}^2-3r^2)/(2(2\tilde{a})^3-6a(2\tilde{a})^2+(2a)^3)$ for $2a \le r \le 2\tilde{a}$ where $\tilde{a}=a(1+\delta)$ such that $\delta=0.1$ is the ratio of the polymer size to the particle radius and sets the range of attraction \cite{varga2015hydrodynamics}. $U_A$ is the strength of attraction potential, and $F_A=U_A/a$ sets the strength of attraction, which can be varied relative to the thermal energy $k_BT$. We use linear springs \cite{luding2008cohesive} to model the contact interactions $\mathbf{F}_C$ and $\mathbf{T}_C$, incorporating both sliding $\mu_S$ and rolling friction $\mu_R$ coefficients. Coulomb’s friction law relates the tangential sliding force $\mathbf{F}_{C,t}$ (proportional to the tangential spring deformation) and the quasi-rolling friction force (proportional to the relative rolling displacement) $\mathbf{F}_{C,r}$ to the normal contact force $\mathbf{F}_{C,n}$ by $|\mathbf{F}_{C,t}| \leq \mu_S |\mathbf{F}_{C,n}|$ and $|\mathbf{F}_{C,r}| \leq \mu_R |\mathbf{F}_{C,n}|$. Further details are available in the Supplemental Materials.

Note that the quasi-rolling force $\mathbf{F}_{C,r}$ does not contribute to the force balance but results in a rolling torque that resists the relative rolling motion between the contacting particles. As will be shown in the subsequent subsections, the rolling torque (and hence, rolling coefficient of friction $\mu_R$) results in the roll-resistant contacts, which have been shown to result in bending rigidity and microscopic aging in colloids \cite{bonacci2020contact, bonacci2022yield}. Thus, rolling friction is the micro-mechanical underpinning for yield stress in highly dilute attractive colloidal dispersion with volume fractions below percolation limits.

\begin{figure}
\includegraphics[width=0.65\textwidth]{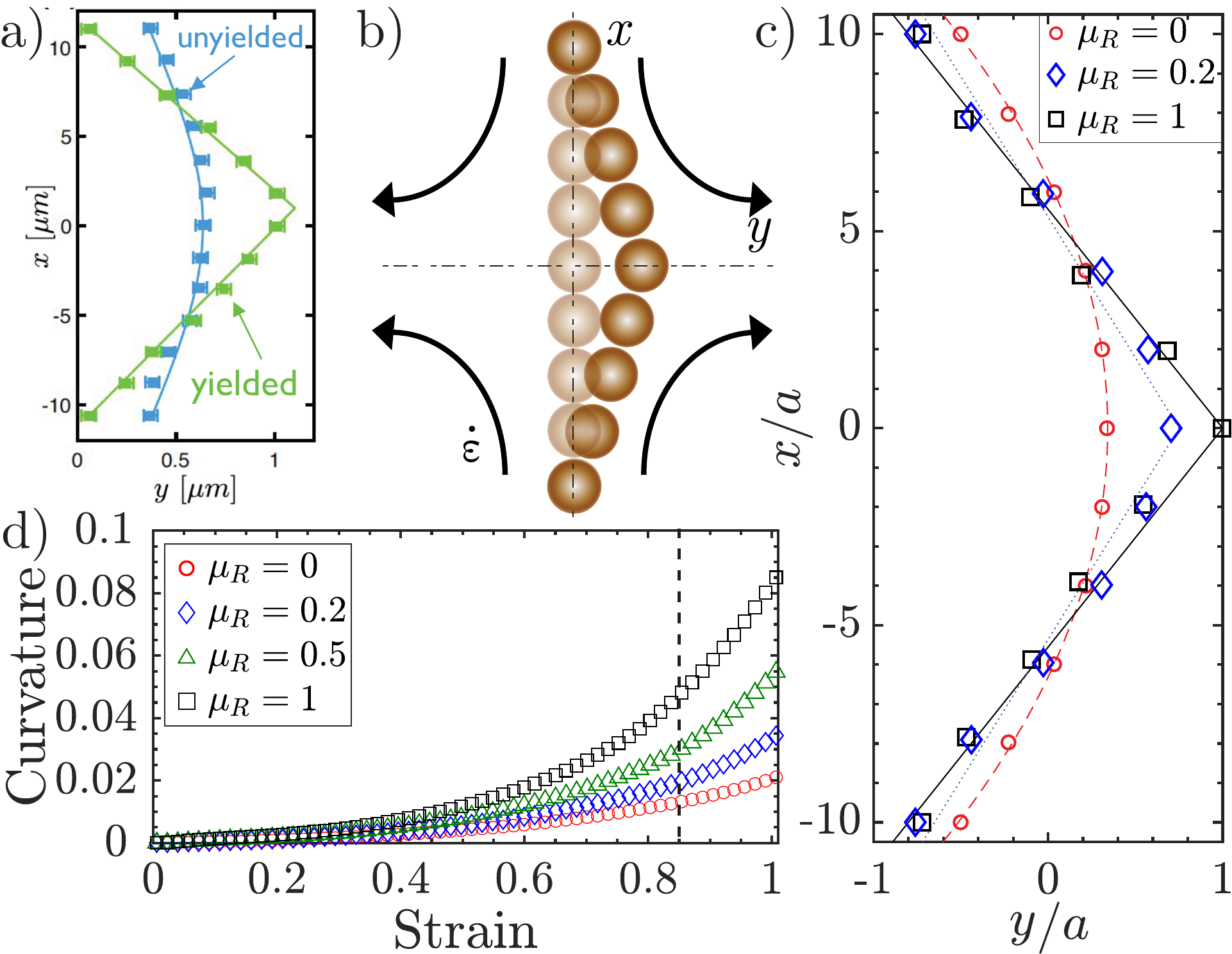}
\caption{\label{fig:fig2} a) Two experimental snapshots of the reconstructed flow of a silica particle rod’s particle positions in the shear (x,y) plane, just before and just after a yielding event. Reproduced with permission from Bonacci \textit{et al.} (2022) \cite{bonacci2022yield}. b) We emulate the experimental conditions in (a) by placing a chain of 11 spherical particles in a planar extensional flow with an imposed rate $\dot{\epsilon}=3 s^{-1}$ (Pe=5.66).  The attraction strength is $U_A=10k_B T$. c) Snapshots of particle positions showing the rod deformation with three different rolling friction coefficients. 
The localization of the deformation and buckling can be quantified by calculating d) the local curvature of the particle rod at its centre obtained by fitting a quadratic polynomial to the positions of the 5 central particles. 
}
\end{figure}

\textit{Simulation details.}---We initialize simulation with all the particles dispersed uniformly in the cubical domain at time $t=0$. Because of their thermal energy, particles undergo Brownian diffusion. They start forming aggregates if the attraction potential well is deep enough ($U_A \gg k_BT$) to kinematically arrest particles upon coming close enough as time progresses. Thus, the structure \textit{builds-up} and in principle, one can quantify it with a dimensionless structure parameter $\lambda$ \cite{larson2019review}, which is 0 at $t=0$ and increases as the system evolves in time to attain a steady state structure with $\lambda = 1$ in a quiescent state. However, applying a finite shear deformation as a constant shear rate $\dot{\gamma}$ or $Pe$ leads to the \textit{destruction} or prevention of structure build-up as it releases mechanical energy in the system. Thus, for a finite $Pe$, the systems attain a value of $\lambda$ between $0 \le \lambda \le 1$ in the steady state depending on the balance between various state variables like $U_A$, $Pe$, and $\phi$. This results in the observed shear-thinning in the steady-state flow curve of attractive colloidal dispersions. At a sufficiently low $Pe$, however, the shear deformation is small enough such that $\lambda \approx 1$ is believed to result in a finite yield stress. However, this is not a sufficient condition. 

\textit{``Roll'' of friction in determining the yield stress.}---Figure~\ref{fig:fig1}a shows the dimensionless steady-state flow curve for attractive colloidal dispersions with the same inter-particle attraction strength $U_A=10k_BT$ to promote particle aggregation, but three different rolling friction $\mu_R$ values. It is the attraction strength that determines the kinetics of aggregation, and hence, structure \textit{builds-up} irrespective of $\mu_R$ as shown by the snapshots of particle positions in the shear plane once a steady state is achieved in Figure~\ref{fig:fig1}I, II, and III at low $Pe$. However, a distinctive plateau in the dimensionless shear stress, i.e., a finite yield stress $\sigma_y$, appears only for finite values of $\mu_R>0$ at low $Pe$. In the absence of rolling friction, i.e., $\mu_R=0$, the low $Pe$ plateau in the steady state shear stress disappears, and the colloidal dispersion flows like a viscous liquid irrespective of the magnitude of the applied deformation and structure build-up. Increasing the rolling friction magnitude results in an increase in the yield stress of the dispersions, as shown in Figure~\ref{fig:fig1}c. 

The total shear stress $\sigma(\dot{\gamma},\lambda(\dot{\gamma}))$ can be split into structure parameter $\lambda$ and shear-rate $\dot{\gamma}$ dependent contributions \cite{massaro2020viscoelastic} such as elastic contribution, $\sigma_y(\lambda)$, viscous contribution, $\eta_{\lambda}\dot{\gamma}$, and a structure independent and rate-independent Einstein viscosity contribution from the colloidal matrix, $\eta_{\lambda=0}=\eta_0(1+2.5\phi)$. Figure~\ref{fig:fig1}b plots these different contributions to the total shear stress as a function of applied shear rate for the $\mu_R=1$ dispersion. 

At low $Pe<0.1$, a constant elastic contribution $\sigma_y(\lambda)$ dominates as a function of the aggregated structure $\lambda$. The structure is fully built up, and hence $\lambda$ is close to its maximum value of 1 at low $Pe$. This elastic contribution is present only for a finite $\mu_R$ and leads to the observed yield stress in Figure~\ref{fig:fig1}a. As explained in the subsequent subsections, this constant elastic contribution arises from the interplay between the constraints on the relative motion between the particles, which restricts the rearrangements of the particles in the aggregated clusters. This leads to a significant resistance to flow at low $Pe$ and, hence, a finite yield stress. In the absence of $\mu_R$, there are no constraints on the rearrangements of particles in the clusters, as a result of which the resistance to the flow and, hence, the elastic contribution diminishes, resulting in a liquid-like flow as shown in the steady state flow curve of Figure~\ref{fig:fig1}a. 

At the intermediate $Pe$ values, $100>Pe>1$, the structure is progressively \textit{destroyed} ($\lambda$ falls below 1) due to the imposed shear deformation. As a result, the structure-dependent elastic contribution reduces and becomes negligible, and the resistance to the flow is increasingly dominated by the viscous contribution in the system, as can be seen from the transition in the dominant contribution to the total stress in Figure~\ref{fig:fig1}b. Finally, at large $Pe$ values ($>100$), the total shear stress in the systems is dominated almost solely by the Einstein viscosity contribution as the large imposed shear deformation destroys the structure, lowering $\lambda$ to $\approx 0$. This destroyed structure at large $Pe$ is shown in Figure~\ref{fig:fig1}IV.

\begin{figure}
\includegraphics[width=0.8\textwidth]{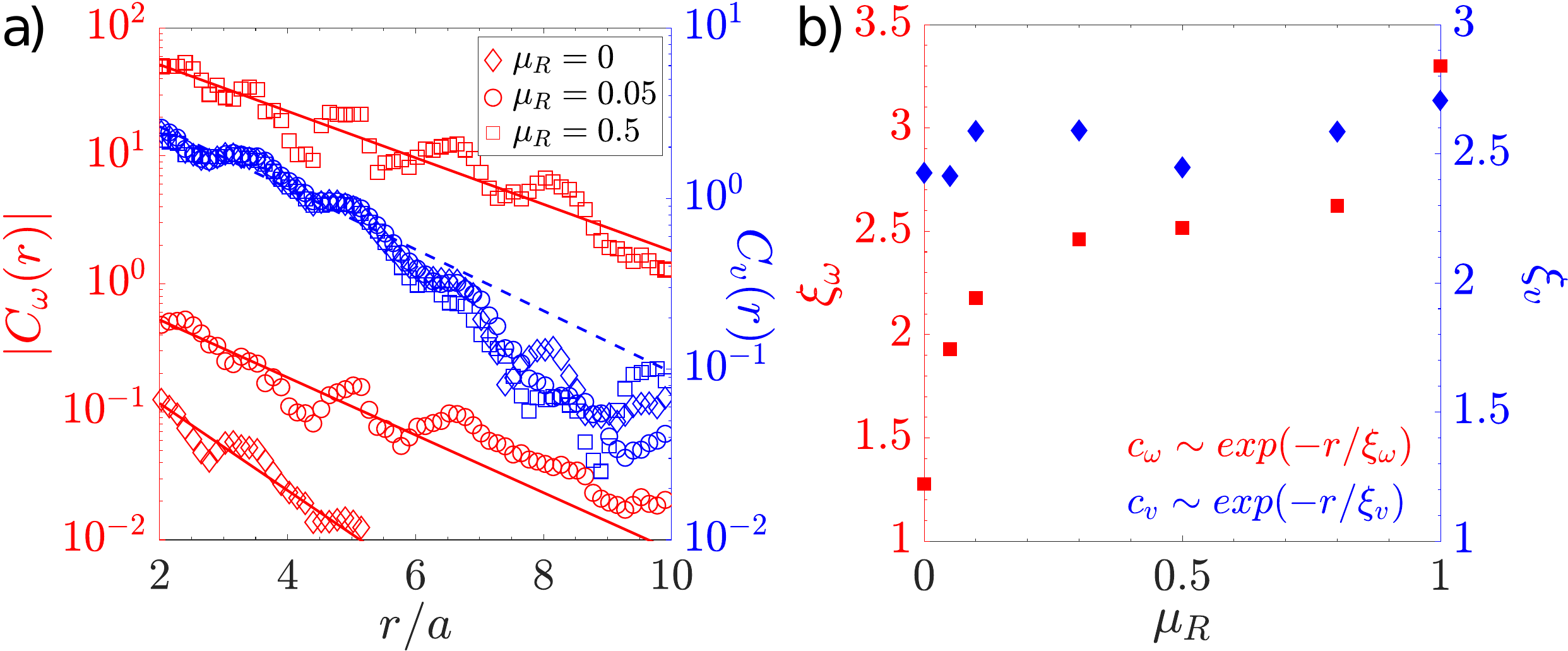}
\caption{\label{fig:fig3} a) Steady-state translational ($v$) and angular ($\omega$) velocity correlations and b) the corresponding correlation lengths $\xi_v$ and $\xi_{\omega}$ for three different rolling friction values for $Pe\approx 0.002$, i.e., in the yield region of the flow curve. 
}
\end{figure}

\textit{Microscopic underpinnings of the yield stress.}---The experimental snapshots in Figure~\ref{fig:fig2}a depict the reconstructed flow of silica particle rods in the shear xy plane, captured just before and after a yielding event. This experiment, conducted by Bonacci et al. (2022) \cite{bonacci2022yield}, involved fixing the end particles while subjecting the middle particle to an increasing force to surpass a critical bending moment such that the rod yields. This experiment demonstrated that solid-solid contacts give rise to a bending stiffness, the evolution of which in time governs the shear-modulus and yield stress aging of \textit{dense} aqueous silica and polymer latex suspensions at moderate ionic strengths \cite{bonacci2020contact,bonacci2022yield}. The pre-yielding positions of the particles (depicted in blue) align closely with the predictions of the Euler-Bernoulli equation. However, post-yielding positions (depicted in green) reveal a distinct structural change, forming two straight segments connected at a finite angle. This deviation from the pre-yielding behavior signifies a significant rearrangement of the particle rod structure following the de-pinning of the contact at the middle of the chain. This suggests particle-particle contacts must support finite torques and resist relative rotational motions. We simulate these experiments to show that a finite $\mu_R$ captures the micro-mechanics of these observations and, hence, is the reason behind the emergence of yield stress at the bulk scale. 

We subject a chain of 11 attractive particles with and without rolling friction to a planar extensional flow on, as illustrated in Figure~\ref{fig:fig2}b, to emulate the experimental conditions described in Figure~\ref{fig:fig2}a. The imposed rate of $\dot{\epsilon}=3 s^{-1}$ (corresponding to a Péclet number of 5.66). The attraction strength was set to $U_A=10k_B T$. 
Our simulation results, as shown in Figure~\ref{fig:fig2}c, reveal that without rolling friction ($\mu_R=0$), the chain does not exhibit yielding behavior as observed in experiments \cite{bonacci2022yield}. However, non-zero values of $\mu_R$ result in Euler-Bernoulli beam-like deformation and subsequent buckling of the rod structure at extensional rates surpassing a critical threshold, mirroring the experimental findings \cite{bonacci2022yield}. Specifically, the chain of spheres deforms into two straight segments connected at a finite angle at the center, akin to the post-yielding configurations observed experimentally due to the de-pinning of the contact from roll resistant contacts \cite{bonacci2020contact, bonacci2022yield}. Figure~\ref{fig:fig2}d depicts the evolution of the local curvature of the particle chain at its center. It reveals an initial gradual increase in curvature as the rod undergoes deformation under the imposed extensional flow, consistent across all values of $\mu_R$. However, beyond a critical strain for non-zero values of rolling friction, the local curvature suddenly increases, which indicates buckling-induced yielding as depicted in Figure~\ref{fig:fig2}a and c. 


\begin{figure}
\hspace{-4mm}\includegraphics[width=0.6\textwidth]{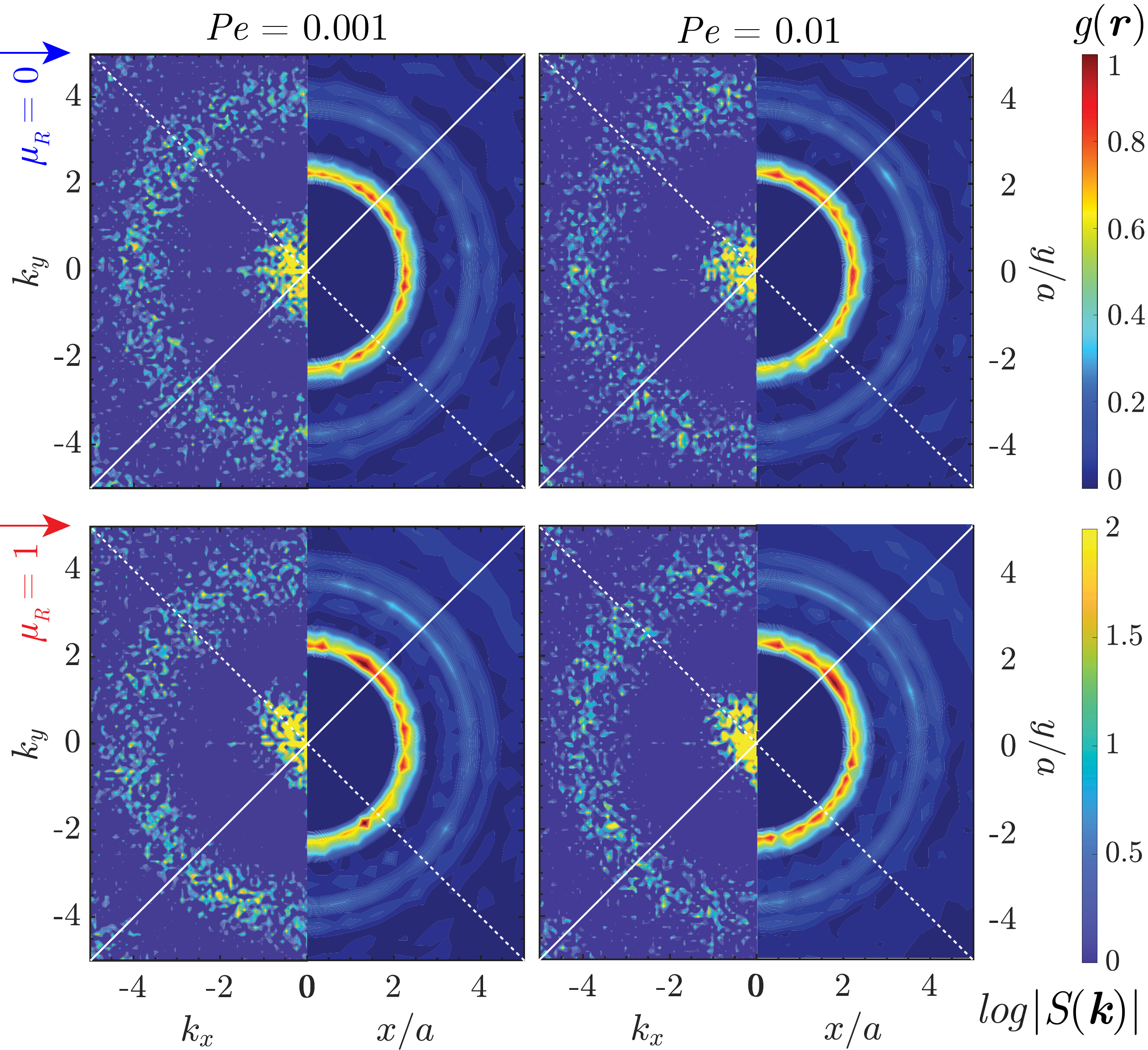}
\caption{\label{fig:fig4}Aggregate structure quantified using the pair distribution function $g(\mathbf{r})$ (right half in each panel) and the structure factor $S(\mathbf{k})$ (left half in each panel) projected in the shear (xy) plane at $Pe$ values in the yielding region of the flow curve and two disparate $\mu_R$. These contours indicate an anisotropy in the structure of the aggregates along the expansion axis of the shear flow (solid line) for a finite $\mu_R$ (bottom row). On the contrary, the structure is isotropic without $\mu_R$ (top row). These signatures can be sought out in experimental structure measurements to elicit the presence of constraints of the relative rolling motion between particles, which results in the yield stress in these systems.}
\end{figure}

\textit{Mesoscopic frustration of particle rotation.}---To quantify the effects of the above microscopic mechanisms at the mesoscale, we calculate the translational $v$ and angular $\omega$ velocity correlations. Figure~\ref{fig:fig3}a depicts the translational velocity correlation $C_v(r)$, and the corresponding correlation length $\xi_v$ does not change much with increasing $\mu_R$. However, a finite $\mu_R$ changes the angular velocity correlation $|C_{\omega}(r)|$ by several orders of magnitude and increases the corresponding correlation length $\xi_{\omega}$. This increase in $|C_{\omega}(r)|$ with $\mu_R$ indicates that the relative rotational motion between the particles in a cluster becomes highly correlated due to the constraints on the relative rotational motion between the particles. This increased angular velocity correlation opposes the particle rotation induced by the imposed simple shear flow. This resistance is present only for a finite rolling friction between the particles, resulting in yield stress. An absence of $\mu_R$ results in no long-ranged correlation between the particle rotational velocities, and hence, the resistance to the shear flow is negligible, resulting in the absence of yield stress.

\textit{Macroscopic signature of microscopic constraints in beam scattering.}---Experimentally tracking particle angular velocities is challenging, though some advances have been made in recent years \cite{niggel2023}. Furthermore, measuring the rolling friction has also been challenging experimentally due to the submicroscopic sizes of the colloidal particles \cite{scherrer2024measuring}. As a result, searching for the signatures of the microscopic rolling resistance at the micro or meso scale is challenging at the current moment. On the contrary, beam scattering has been popular and comparatively accessible, and it has been widely used to characterize the microstructure in colloids and suspensions. Analyzing the scattering pattern of dilute attractive colloidal dispersions should identify the macroscopic signatures of the microscopic rolling resistance and mesoscopic frustration of particle rotations as changes in the scattering patterns of dispersions with and without yield stress.

We calculate the beam scattering pattern in the form of the structure factor $S(\mathbf{k})$, which is essentially the Fourier transform of the particle pair distribution function $g(\mathbf{r})$ projected in the shear xy plane. These are depicted in Figure~\ref{fig:fig4} in the low $Pe$ regime for dispersions with (bottom row) and without (top row) the presence of microscopic rolling friction and, hence, bulk scale yield stress, respectively. The qualitative subtle differences in the particle cluster snapshots in Figure~\ref{fig:fig1} at finite $\mu_R$ can be seen as an anisotropy of the particle structure at domain length scale in the extensional direction (indicated by the solid line in our figures) of the imposed shear flow. Due to the resistance to the rolling of the frictional particles, they cannot freely organize in spatially uniform clusters compared to non-frictional particles. As a result of the flexural rigidity at the microscale and frustrated rotation at the mesoscale, the particle clusters can only get stretched along the extensional axis of the imposed shear flow. This results in anisotropy in the beam scattering patterns at finite $\mu_R$. 



In contrast, when $\mu_R$ is zero, the structure of the aggregates remains isotropic regardless of the shear conditions. This isotropy indicates that particles can freely rearrange in response to shear forces without the constraints imposed by rolling friction, leading to a more uniform distribution without preferred orientation. The absence of rolling friction allows for greater fluidity within the suspension, preventing the formation of anisotropic structures otherwise stabilized by finite $\mu_R$. These structural signatures—anisotropy with finite $\mu_R$ and isotropy without it—offer a pathway to experimentally detect and quantify the constraints of relative rolling motion between particles.

Experimental techniques such as confocal microscopy and scattering experiments (e.g., small-angle neutron scattering or X-ray scattering) can be employed to measure $g(\mathbf{r})$ and $S(\mathbf{k})$ in sheared suspensions. By analyzing these measurements, one can potentially infer the presence and magnitude of rolling friction, which directly correlates with the observed yield stress in these systems. The structure anisotropy induced by finite $\mu_R$ points to significant micromechanical constraints that impact the yield stress and overall rheological behavior of colloidal suspensions. Future experimental work should focus on detailed structure measurements, especially in the shear plane, to validate these theoretical predictions and further elucidate the mechanisms governing aggregate formation and stability in complex fluids.

\textit{Conclusions.}---In conclusion, our study offers valuable insights into the micromechanical origins of yield stress in soft particulate gels, specifically focusing on model dilute depletion gels \cite{dullaert2005model}. Through coarse-grained simulations, we have identified the crucial role of microscopic constraints on the relative rotational motion between bonded particles in affecting the kinetic of particles at the mesoscale and, consequently, the structure and rheology at the macro-scale. In particular, we show evidence that attraction-induced structure formation alone is insufficient, and microscopic constraints on the relative motion between particles are necessary for the emergence of yield stress in dilute depletion gels. Further exploration of these micromechanical principles promises to unlock new avenues for developing tailored soft particulate gels with enhanced performance and functionality, thereby addressing the evolving needs of diverse industrial and biomedical applications.

\bibliography{bib}

\end{document}